# 61 ways to measure the height of a building: an introduction to experimental practices

F. Bouquet[1], J. Bobroff[1], A. Kolli[1] and G. Organtini[2]

[1] Université Paris-Saclay, CNRS, Laboratoire de Physique des Solides, 91405, Orsay, France.
[2] Sapienza Università di Roma & INFN Sez. di Roma, Italy

E-mail: frederic.bouquet@universite-paris-saclay.fr

**Abstract**

We created an introductory physics activity for undergrad students consisting in measuring by different methods the same physical quantity. It allows to confront students with questions of uncertainty, precision, and model versus theory. The aim was to measure the height of a building using only a smartphone and everyday low-cost equipment. We designed 61 methods to do so and tested most of them. These methods can be implemented in various pedagogical scenarios, engaging students into a concrete task, outside of the lab, easily set up at almost zero cost.

# 1. Introduction

Experimental physics is an important part of physics learning in order to teach students how to think "like a physicist". [1] We developed a specific teaching to get undergraduate students face three basic questions:

- How to design and build an experimental setup?
- What is the precision of the measure, and why it is a key issue?
- How is it possible that an experimental result does sometimes not follow the model's predictions?

We wanted these questions to come naturally to students, not as an academic request from their teachers. We therefore asked students to design and perform measurements of a given unique quantity with different methods, to compare the various results and to discuss them within theoretical frameworks. We decided to use smartphones because they give students a familiar common tool they can use to do physics experiments in full autonomy. It has been widely demonstrated that smartphones can be used to perform numerous physics experiments [2, 3, 4, 5] since they contain many sensors [6] whose data can be accessed with easy-to-use applications. Smartphones also have the advantage to reduce the cost of an experiment, to engage students, and to allow distance learning, especially in present times of possible lockdowns. [7]

We based our teaching on the famous urban legend of Niels Bohr and the barometer. [8] When asked how to measure the height of a building with a barometer, a student — young Niels Bohr — invents a dozen or so experiments that do respond to the question avoiding the solution expected by the teacher. We revisited the barometer question as: "How many different ways are there to measure the height of a building with a smartphone?" Unlike Bohr's legend, we further asked our students to carry out the experiments they thought of and to evaluate how the results compare with one another. It turns out that smartphones allow to perform many more experiments covering many fields: mechanics, magnetism, optics, ...

The aim of this article is to show how, using only low-cost equipment, this challenge lets students face directly the questions of good experimental practices. We first present various methods to answer this smartphone question, and then discuss ways to use these experiments in specific teaching contexts and how this could renew the usual approach to labs in undergraduate curricula.

# 2. Converting Bohr's legend into actual measurements

We translated the Bohr's legend into simple constraints: measuring the height of a building several meters high using only every-day objects and a smartphone (with its internal sensors: accelerometer, gyroscope, light sensor, magnetometer, GPS, barometer, microphone and camera). We mainly use the "phyphox" app [5] to access the sensors measurements, because it is easy to use and is available in many languages, but other apps would also work as well [e.g., physicsToolbox].

To test the possibilities of this challenge we found up to 61 methods. [9] Table 1 presents them, as well as the results obtained carrying out 46 out of the 61 methods using our laboratory building as a test sample (see Figure 1); the other methods (labelled with a star) were not tested because of lack of material or technical limitations. Most of these methods have already been described elsewhere, albeit not always on the scale of a building. They are fully described in the appendix of this article.

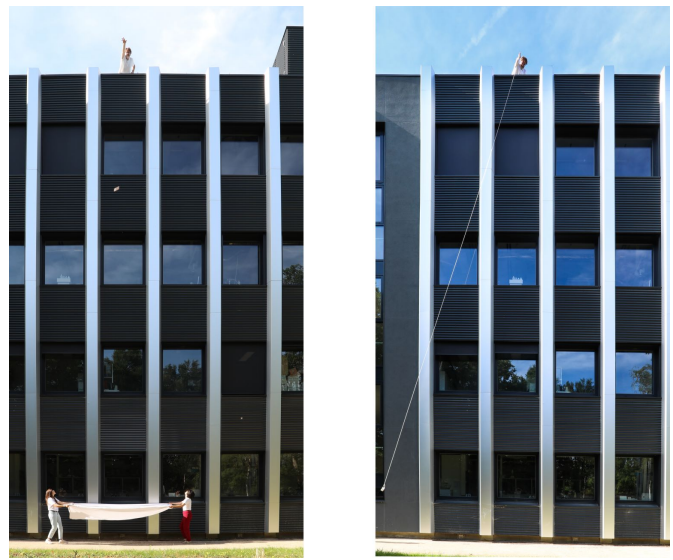

Figure 1: Measurement of the height of the building using method #1 (free fall of a smartphone) and #13 (giant pendulum, using the smartphone's gyroscope). Results are presented in Table 1.

As seen in Table 1, these methods span a large spectrum of physics domains: free fall mechanics (#1 to #9), pendulum physics (#10 to #20), trigonometry (#21 to #26), picture analysis and optics (#27 to #30), acoustics (#39 to #43), wave physics (#44 to #48, #55 and #56), inverse squared law (#49 to #54), even general relativity (#60), and, of course, using the barometer that some smartphones possess as expected by Bohr's physics teacher (#36). Keeping the spirit of Bohr's legend, some very simple methods use the smartphone in an unorthodox but efficient way, such as counting stair steps (#35) or as a scale to measure the height (directly #34, or through its shadow #21).

A striking feature of Table 1 is the dispersion of the results, even though they were performed on the same 15-m building,

as illustrated by Figure 2. One can classify the methods among different cases:

**Some methods show a reasonably small error bar encompassing the 15-m value**, as expected for a well-executed measure, and reassuringly corresponding to many of the results. Any error bar smaller than 0.5 meter represents better than 3% of precision, such as the giant pendulum experiments (#13 to #16), picture analysis (#26 to #30), pressure measurement (#36), and some simple techniques (#34, #35).

**Other methods feature a large error bar encompassing the 15-m value**, indicating a very inaccurate measurement. This can be due to a measure that in itself is imprecise, such as using the GPS (#38), or the WIFI signal strength (#51), or a method that will be very sensitive to small uncertainties and amplify them, such as some trigonometry method (#25).

**Another group displays small error bar but not encompassing the 15-m value**, which indicates that something is off. It may be that the error bars were just underestimated, which could be the case for example for the measure of the smallest building shadow length at the equinox (#23). Another important fact to consider is whether the hypotheses underlying the model used to derive the result are verified. For example, in the case of the shadow length, one assumes that the ground is perfectly horizontal. The free fall methods give another striking example. Most of these methods assume that air friction can be neglected, which is not the case when using a tennis ball falling over 15 meters, as was the case for the results presented in Table 1 and Figure 2. Methods based on the timing of the fall of an object from the top of the building (#2, #3, #4) result in an overestimation of the height (an increase of the fall time), methods using the end velocity of a falling object (#5, #6, #8) result in an underestimation of the height (a decrease of the velocity). Using a method that does not assume that air friction is negligible, such as letting a smartphone fall from the top of the building with its accelerometer on (#1) will give a correct result (see Figure 1 for the setup, and appendix A for the data).

**A last group of experiments features gigantic error bars, several orders of magnitude higher than the actual height.** This indicates that the method is not suited for the task at hands. It is the case when a method could work in a very idealised world but would require unreasonable precise measurements. For example, general relativity (#60) implies that time is not the same at the top of the building and at the bottom, since the strength of Earth gravity g will depend on the altitude. Assuming that the minimum difference between two smartphone stopwatches running for one hour, one at the top, one at the bottom of the building, is 1 ms, leads to a minimum height that can be measure of 3000000 km (see appendix A).

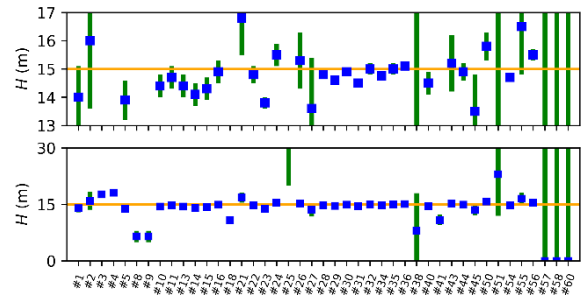

Figure 2: results of the different methods presented in Table 1. The orange horizontal line represents the value 15.0 m. The two panels present the same data, on a different scale.

## 3. Using Bohr's legend to create engaging teachings

We first tested this challenge as a complete open question where students had to invent and build their own way to measure the building's height. It allows for a complete student autonomy, but the proposed solutions were almost always the same, mainly using free fall timing, atmospheric pressure measurements, and pictures of the building with a scale. To broaden the diversity of techniques we now give our students some of the 61 methods presented above to choose from. We choose each time a subset of these methods that fits our pedagogical objectives and the available time and material.

For example, in a 1.5-hour session, we asked 20 second-year university students to compare 10 different methods that cover different fields of physics and could be performed rapidly (#2, #3, #4, #9, #10, #12, #14, #15, #36, #38, #39). Students then had to decide which method was the most accurate and why. They not only had to perform as many experiments as possible, but they also had to work in groups and discuss altogether what "most precise" meant and how to assess this quality within the time constraints. We also used a similar pedagogical scenario with high-school physics teachers in half-day lifelong training. Here we chose a wider span of experiments focusing on physics phenomena, in order to demonstrate the possibilities of smartphones as a measuring device. Teachers could test several setups and use them afterward with their own pupils. Other subsets can be used for different objectives, working on uncertainties for example, or thematic subsets, such as free fall, pendulum or acoustics. [9]

In each of these frameworks, students clearly enjoyed the opportunity of getting out of the lab and doing measurements that had a direct and concrete meaning to their daily life. The gigantic size of the "sample" is an engaging challenge. We found out that giving them access to a wide variety of low-cost objects (cords, tubes, tapes, foam, papers, straws, balls, balloons, playdough …) and letting them tinker their set-up, challenging their creativity, was effective. Taking the time to organise a collective comparison and a debriefing session after the measurements is a key point.

When the top of the building is not accessible, some adjustments can be made, either by measuring the height of an accessible window or indoor balcony; if going out of the classroom is not an option, measuring the height of the room is also a possibility, albeit mundane.

Undergraduate labs are usually devoted to measure a specific law with a given setup. Here, the variety of available methods allows a new approach by comparing different experiments and models. This approach is particularly suited for an introductory level, since the physics at play is often easily accessible. The experimental setups that the students must build are not complex, but trigger questions of protocol and precision: how to make the measurement, how many times? Letting students compare their different results with each other forces them to address the issue of uncertainty in a more meaningful way than during a traditional students' lab or theoretical course. It forces students to question the quality of their experimental design, not just taking it for granted as in a classical lab. They also have to cross-examine the hypothesis behind the models.

## 4. Conclusion

Open-ended activities in students' laboratory activities have been shown to be more efficient than guided activities to develop a more expert-like behaviour toward experimental physics. [10] We propose here the building's height question as an effective way to engage students to invent their own experimental setups and confront their results.

This classroom activity can also be easily adapted to other publics. For example, we used it in a science museum for a general public outreach activity. Last but not least, this "smartphone physics challenge" can be implemented for remote teaching, where students carry out experiments at home. This could serve for homework activities or as distant labs sessions during lock-down periods such as the brutal ones experienced by many of us recently.

## Acknowledgements

We gratefully acknowledge Ulysse Delabre and Joël Chevrier for discussions and inspiration, and Anna Kazhina for her work on the graphic presentation of the methods. We thank the students who participated to these projects, K.V. Pham for welcoming this new teaching in the physics curriculum of Paris-Saclay University, and the phyphox team for their helpfulness. We thank the Institut Villebon - *Georges Charpak* for helping us to develop new teachings. This work has been partially supported by the Erasmus+ teaching mobility program. It also benefited from the support of the Chair "Physics Reimagined" led by Paris-Saclay University and sponsored by AIR LIQUIDE, and also from a grant "pédagogie innovante" from IDEX Paris-Saclay.

## Appendix A. Technical Details

We present here technical details on the principle and realisations of the 61 methods presented in Table 1. These methods have also been graphically summarized, for distribution among students.[9]

### A.1. The teacher's solution

In Bohr's legend, the teacher expects a specific solution: the barometer is used to measure the variation of atmospheric pressure $\Delta P$ between the top and bottom of the building. The altitude $H$ of the building follows by $H = \rho g \, \Delta P$, with ρ, the density of air, close to $1.2$ kg/m$^3$. This method can be performed with smartphones that are equipped with a barometer sensor. [11] For a better measurement, a calibration of both the barometer and ρ can be performed by measuring the difference of pressure between two points of known altitude, such as the distance between one's head and feet. As seen on Table 1, this method can be quite precise.

### A.2. Methods using free fall

When the air friction is neglected, the fall time $t$ of an object with zero initial velocity gives the height[2] [12] through $H = \frac{1}{2} g t^2$. Caveat: throwing objects from the top of a building is potentially hazardous; we used a tennis ball to attenuate the risks.

The fall time of the tennis ball can be measured by different means: timed with a stopwatch app (#2, not a precise method), video-analysed (#3), or using an audio recording (#4). For the latter, a neat method to produce a sound at the beginning of the fall is to tie the ball to a balloon and pop it. The audio analysis can be performed after the measurement on a PC using software like Audacity or on the fly with some smartphone apps such as phyphox' acoustic stopwatch. [13] The audio analysis leads to a slightly better resolution than the video because of its higher sampling rate.

A close variant consists of measuring the velocity $v_f$ of the falling object at impact, using $H = v_f^2/(2g)$. The easiest way to measure $v_f$ is to film the end of the fall and analyse the video frame by frame (#5). A more sophisticated way to measure this velocity would be to audio record the fall and perform a Doppler analysis of the sound, assuming the falling object emits a continuous sound (a Bluetooth speaker for example, #6*). [14] [15]

Even though these methods are based on the same mechanical model (free fall without air friction), they give different results: timing the fall gives larger than expected results (for #4, $H = 18.1 \pm 0.3$ m, see Table 1), whereas measuring the impact velocity gives a much lower result (#5: $H = 13.9 \pm 0.7$ m). This can be explained by the effect of air friction, which reduces the impact velocity, making the building appear smaller in the latter method, or increases the fall time, making it appear larger in the former.

Letting the smartphone free fall itself is a way to take air friction into account (#1): the accelerometer records the deceleration due to friction. Integrating twice the signal [16] [17] gives $H$. One can use a bedsheet hold by two people to catch the smartphone safely, like fireman life net. However, one should worry especially of unexpected winds. When we threw our smartphone from the 15-m roof of the building, it rolled and looped during the fall, which explains the not-monotonous acceleration curve we obtained (see Figure A1). Bracketing the value of acceleration by linear curves gives $H = 14.0 \pm 1.1$ m. Since the effect of air friction is taken in account by this method, providing a smoother fall (with a parachute for example) could lead to smaller uncertainties.

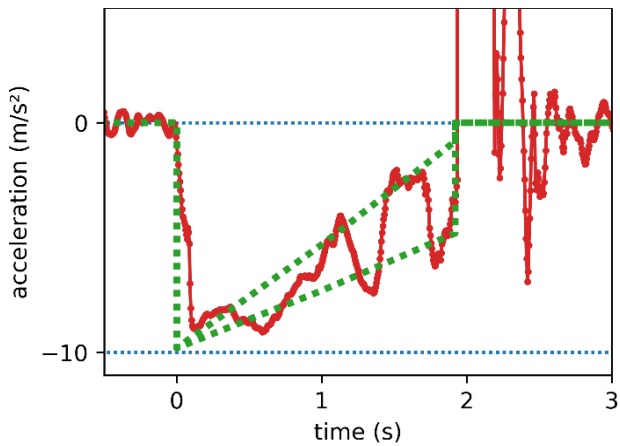

Figure A1: Acceleration recorded by a smartphone during its fall from the top of the building (see Figure 1). Gravity was subtracted from the accelerometer raw data. The huge spikes after 2 s correspond to the landing and oscillations of the smartphone in the life net afterwards. The beginning of the fall defines $t = 0$. The fall time $1.92 \pm 0.03$ s leads to $H = 18.1 \pm 0.6$ m neglecting air friction. The effect of air friction is shown by an acceleration smaller than $g$ and tends to zero during the fall, even though the erratic rotation of the falling smartphone makes the curve non-monotonic. The dashed lines are the bracketing curves used to determine $H$ by integrating twice, leading to $H = 14.0 \pm 1.1$ m.

Other methods using free fall can be used. A horizontal initial velocity $v_i$ can be added to the ball when throwing it from the top of the building, and measuring $v_i$ using video analysis and the horizontal distance $d$ the ball reached at impact leads to $H$ (#7*). One can also let the ball fall and time the rebounds, [12] [18] either through audio (#9) or video (#10) analysis. The time between two rebound leads to the coefficient of restitution, and –- assuming this coefficient is constant –- to the height of the first fall. All these methods assume that air friction is negligible, which we know was not the case in our experiments with a tennis ball. Using a smaller and heavier ball, such as a golf ball, might have yield to different results.

*A.3. Methods using a giant pendulum*

Since Galileo, pendulums of known length have been used to measure time. Building a giant pendulum the size of the building and timing its period $T$ leads to $H = g(T/2\pi)^2$. Some care should be given to the construction of the pendulum, so that it swings nicely and does not rotate in every direction. The swing construction, with two wires separated by a gap helps a lot.[3]

To measure the period, the simplest way is to use the smartphone stopwatch (#10). Any sensors can be used: video analysis[19] (#11), accelerometer [3] [19] (#12), gyroscope [19] (#13), magnetometer [20] (#14, either using a permanent magnet as the weight, or hanging the smartphone itself to the pendulum and measuring the Earth magnetic field), light sensor [21] (#15), proximity sensor (#16), audio record (#17*) using a Bluetooth speaker and looking for the audio modulation due to the distance to the source varying.

Using the movement sensors (accelerometer and gyroscope) to measure the period of a pendulum has been well reported in the literature. [3] [22] It has the advantages of measuring the oscillation amplitudes (allowing for an amplitude decay study), whereas other sensors detect only the period through the passage of the pendulum. On a giant pendulum the oscillation angle gets smaller and the period increases, making the signal smaller and more difficult to detect: on our 15-m pendulum, the signal/noise ratio is only about 5 for the gyroscope data and the signal for the accelerometer is too small to be observed.

The giant pendulum setup can also be used to measure the height of the building using laws of a body in rotation: the centripetal acceleration $a_c$ (#19*) and the velocity $v$ of the pendulum (#20*) are related to the angular velocity ω through simple equations. [23] The accelerometer and gyroscope of a well oriented smartphone can measure $a_c$ and ω, [19] and the velocity could be measured by either Doppler effect [14] [15] or beats between two speakers, [24] one swinging with the pendulum, one motionless on the ground. In practice these methods are difficult to implement on a setup the size of a building.

A variant (#18) is to measure the period of a giant torsion pendulum. This experiment is more akin a spring experiment than a pendulum one, and the torsion coefficient needs to be calibrated by measuring the period of the pendulum for a known length of wire. This method supposes that the torsion torque is equally spread along the wire and that the connections to the wire are perfect, both approximations that require care to achieve. The discrepancy of our measurement $H = 10.8 \pm 0.5$ m to the expected 15-m height is likely due to these issues.

*A.4. Methods using trigonometry*

In surveying, triangulation is a well-known technique to measure distances. [25] Since the smartphone can measure angles with the vertical using its accelerometer, different setups can be imagined using this principle.

Facing the building, it is either possible to determine the angle to the top of the building knowing the distance to the building (#25, $H = 15.5 \pm 0.4$ m), [26] or the angles to the top and to the bottom (#24, $H = 40 \pm 20$ m). To improve precision, it is best to attach the smartphone to a tube, and use the latter as homemade theodolite sight. The former method yields better results if one is standing not too close and not too far from the building (a distance corresponding roughly to the height of the building is good). The latter method, if performed from one's height, standing on the ground, leads to a large uncertainty on $H$ because the result will be highly sensitive to the measure of the angle to the bottom.

A variant is to measure the apparent angle of an object of known size lying on the ground below from the top of the building (#26). The precision of this method will be better if the size of the chosen object is approximatively the size of the building.

Using shadows is also a well-known technique to measure the height of a building using trigonometry since Thales' times. Measuring the shadow size of the building and that of your smartphone gives the ratio of building height / smartphone dimension (#21). A more direct way using the building's shadow is to know the sun elevation in the sky. A digital method uses the phone's GPS to get longitude, latitude and time, and then visit an astronomical website that provides elevation from these data [27] (#22). A more hands-on approach is to use a time-lapse to determine the minimum size of the shadow on either an equinox or a solstice day (#23). On these particular days, at noon, the elevation of the sun is directly related to the latitude. Getting a minimum size from a time lapse is not easy, and needs a proper scale and a good setup for the camera, but the process involves a better understanding of the relative positions of the Sun, the experimenter and the planet.

*A.5. Methods using photography*

The most straightforward method is to take a picture of the building with an object of known size on the image, playing the role of a scale (#28). Care must be taken to perspective deformations. The phone should be kept parallel to the building, and additional software corrections can help improve parallelism even more.

Using laws of geometrical optics, a picture of the building taken at a known distance on a smartphone of known focal length and sensor size (#29) also leads to the height. [28]

Variants of the previous methods can be envisaged by standing at the top of the building and taking a picture of an object of known size on the ground below, either knowing the focal length and sensor size of your phone camera (#30), or measuring the camera angle of view (#27). An estimate of the latter can be determined experimentally with a protractor.

*A.6. Methods using speed of sound*

A direct method is to record the burst of a balloon at the top of the building, and waiting for the ground echo (#42). However, this requires some ideal building configuration for having a chance to work, and we couldn't catch the echo in our case.

In method #39, a balloon is burst at the top of the building, triggering an acoustic stopwatch of both smartphones, albeit with a delay for the bottom one. A second balloon is then burst at the bottom, triggering the stopwatches off, albeit with a delay for the top one. The differences in the stopwatches record is twice this delay, which corresponds to the time the sound travelled the building height. Acoustic stopwatch is available in phyphox app, it is triggered on and off by a sound threshold. [13] If phyphox is not available, or if the noise conditions are difficult, audio-record analysis from both smartphones will yield the same information (#40). Any sharp noise can be used to trigger the stopwatches. Bursting balloons is quite engaging for students, but banging together two pieces of wood also works well.

In method #43, filming in slow-motion from the top of the building the burst of a balloon at the bottom can clearly differentiate the time of arrival of the light (the image of the burst) and sound (the slow motion capture must also record sound, which unfortunately is not the case on all smartphones, especially at higher framerates). Using a 340 fps record we estimated the difference of 15 frames between image and sound (see Figure A2), which yields to $H = 15.2 \pm 1.0$ m.

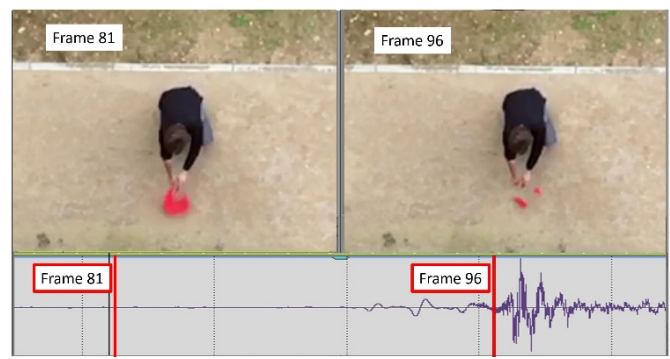

Figure A2: Video analysis of a slow-motion movie taken from the top of the building, at 340 frames per second. Frames 81 and 96 are represented in the two top panels; the balloon is burst around frame 81, and the sound reaches the camera on frame 96 (the sound track is represented in the bottom panel, the vertical red lines correspond to the frames represented above).

Using a phone call to time the travel of sound is tempting, since the electromagnetic waves that carry phone conversations are alike light. A person at the top of the building phones to someone at the bottom, and bursts a balloon. The person at the bottom records two bursts, one having travelled through air and one having being carried out by the communication cell tower (#41). However, additional delays due to the electronic handling of the phone call management need to be calibrated: by popping a balloon when the two phones are close, we found between 260 and 270 ms of electronic delay, much larger that the time needed by an airwave to propagate on 15 m! Assuming this electronic delay constant and performing the experiment gave $H = 10.9 \pm 1.4$ m, which shows that this hypothesis needs to be revised.

In all these methods, the speed of sound can be assumed constant since temperature and humidity variations are neglectible with respect to other sources of error (the temperature dependence is typically less than 0.2% per K). A more general problem is that the saturation of the audio records deforms the audio wave differently with the distance between the balloon and the phone, which introduces some uncertainty. Also, ambient noise and noise reflections will obviously perturb measurements.

*A.7. Methods physics of waves*

Using the phase of sound provides additional methods. The measure of a phase difference is easily converted in distance using the speed of sound (the difference between phase velocity and group velocity is not relevant here). Two audio records are needed to determine a phase difference, and such measurements require more time, care, and analysis skills than previous methods. We performed all sound-phase analysis using Audacity software.[29]

A direct setup is to use two smartphones and a speaker emitting a pure continuous tone. At the beginning, they all are at the bottom of the building, at the foot of an external stair. Both smartphones are audio recording, one is left on the ground, the other is slowly brought up to the top using the external stairs, still recording the continuous tone. Audio analysis of both audio records will show an increase of the phase difference between the two smartphones, [30] related to the distance from the top smartphone to the ground (#44). A higher frequency leads to a lower wavelength which renders the measure more difficult: from a practical point of view, we found best to avoid working above 350 Hz ( $\approx 1$ m wavelength). Working at 200 Hz, we obtained $H = 14.9 \pm 0.3$ m.

If no external stairs exist, a speaker can be used to emit a continuous tone at the bottom of the building, and two smartphones record the audio signal, one at top, one at bottom (#45). The variation of the phase difference $\delta\Phi$ between the two recordings when the frequency $f$ of the tone is changed is related to $H$ and the speed of sound $v_s$ through the relation:

$$\frac{d(\delta\Phi)}{df} = \frac{2\pi H}{v_s}$$

Plotting $\delta\Phi$ as a function of $f$ and determining the slope leads to $H$, though with larger errors than the previous method (we obtained $H = 13.5 \pm 1.3$ m).

A variant of this setup is to keep the speaker at bottom, and have both smartphones at top, one at the vertical of the speaker, the other at a lateral distance $d$ from it (#46*). The relevant relation becomes:

$$\frac{d\delta\Phi}{d(d^2)} = \frac{\pi f}{H v_s}$$

Another type of classical experiments is the standing wave experiments. [31] If one has a long tube running along the building façade, the determination of resonance frequencies should give $H$ (#48*). We did not have such a setup, but presumably a garbage chute for construction could be adequate.

Interferences are another way to determine the height of the building. A double-slit like experiment is possible with two speakers emitting a single continuous tone at a given frequency [32] at bottom, separated laterally by a distance $l$. To ensure the phase coherence, both speakers should be driven by the same sound generator (a smartphone with a split jack connection for example). The sound intensity is measured at the top of the building. The distance $d$ between the first minimums can then be found when moving laterally (#47*). For $(l, d) \ll H$, the equation simplifies into $H = ldf/v_s$.

Light wave can also be used instead of sound waves: diffraction pattern of hair lighted by a laser is well known, [33] and the resulting pattern depends on the size between the screen (the ground at the bottom of the building) and the hair (at the top of the building). It will also depend on the diameter of the hair which can be determined using a drop of water on the smartphone lens to increase the magnification of the camera [34] (#55). The screen of the smartphone can also be a good diffracting object, depending on its technology (#56). The pixels of the screen act as a reflexive diffraction grating; [35] [36] the distance between two pixels can either be obtained similarly than the hair diameter, or just by knowing the number of pixels and the dimension of your screen. Note that great care should be observed with laser. These experiments should be realized during the night, since only unreasonably unsafe lasers would give enough light for the diffraction pattern to be visible otherwise. Using lower power lasers leads to safer but poorer pictures, but we found that software enhancement of pictures can be enough to perform the measurement over 15 m, especially if it is taken with a tripod and a long exposure.

*A.8. Methods based on the inverse-square Law*

The inverse-square law happens when a quantity is freely propagating from a punctual source without any other effect (diffusion, absorption, reflection, interferences ...). After proper calibration, a measure of this quantity can be used to determine the distance to the source. Having the source at one point of the building and the measure at the other allows to determine $H$. Several quantities can be used, with varying degree of precision.

Light is a prime quantity for this method, and it works well (#50): the light sensors in the smartphones are generally good enough to provide accurate quantitative measurements. [37] Care should be taken to the ambient light (night work is necessary). Also important is the orientation of the smartphone and of the light source, since they affect the measurement. We obtained $H = 15.8 \pm 0.5$ m.

Sound also follows the inverse-square law (#49*), but sound intensity measurement made by a smartphone are generally not precise enough, presumably due to the quality of the smartphone microphones or multiple echoes.

Using dedicated apps, smartphones can also measure the power of the Wifi signal emitted by a hotspot. In an idealized world the Wifi power should follow the inverse-square law (#51) but in practice, when trying to calibrate the signal we found that this is not the case. Many reasons could explain this, from numerous reflections of the signal to artefacts from the signal processing, and the low precision of the measure. This method gave the height of our lab somewhere between 12 and 35 m…

When working with a fixed camera setting, the number of pixels occupied by an object on a picture also follow the inverse-squared law (#54). This experiment is easy to perform with a smartphone and gives good results (in our case, $H = 14.7 \pm 0.6$ m using a roughly 1-meter square panel as a target.

The magnetic field generated by a magnet does not follow the inverse-squared law (since this is not a propagation effect); nevertheless, the dependence of magnetic field variation with distance is known ($B \propto 1/r^3$), and since some smartphone can measure a magnetic field the same process of calibration/measurement could be done (#52*). However, for the magnetic field to be detectable over such a distance, large unsafe magnets should be used, and since our lab façade is covered with metal rods we deemed preferable not to test it.

Going down the unsafe road, it was reported that smartphones could be used as poor-man Geiger counters. [38] Since nuclear radiation follows the inverse-squared law having a source of nuclear material should theoretically allow to measure the height of a building (#53*). We did not pursue this avenue any further (for a safer approach see Ref [39]).

*A.9. Direct methods*

Some methods rely on more direct approach rather than specific physics laws, generally in a simpler (but not always precise) way.

Going up the stairs and counting how many smartphones should be piled up to reach the top is an easy and relatively precise way of determining $H$, if done with care on a convenient stairway (#34, using two identical smartphones to alternate them one above the other helps). Using the accelerometer to count the number of stairs and multiplying by the size of a stair also works relatively well (#35). Using a rope weighed by a smartphone and letting it slide down the façade of the building from the top will give a rope of length $H$, which can then be measured with a meter (#31). The latter solution can be made a bit more technical if a pulley is installed at the top of the building to let the rope slide: attaching a smartphone to the pulley with the gyroscope on will record the number of turns the pulley does, [40] which can easily be converted into a distance (#32). Also, if the smartphone attached to the rope has its accelerometer on, the data after two integrations will also give a distance (#33*), in a very similar way of the free fall experiment, or the elevator experiment [16] (#37). These three experiments use the same physics principle, but the smoothness of the elevator generally yields to better results.

Perhaps the most straightforward approach is to use the altitude measure from the smartphone GPS (#38). However this method is very inaccurate since altitude is not what is measured best by a GPS, where a typical 6–8-meter uncertainty is common. [41]

Finally, the most efficient method is to phone the architect and ask him or her how tall is the building.

*A.10. Methods that only work in theory*

Keeping par with the Bohr's legend, it is interesting to explore methods that would only work in an ideal world of "spherical cows". These methods only work in an idealized world where no perturbations are present, and require very precise measurements. Smartphones are obviously not the right tool, but some error bars can be calculated by estimating how tall the building would need to be for a signal to be measurable.

For example, assuming Earth is a perfect magnetic dipole, using the magnetometer to measure the magnetic field at the top and bottom of the building should lead to $H$.

A 15-m change of altitude would correspond to a 0.0008% change of magnetic field, below the standard smartphone sensor resolution. But Earth magnetic field is not exactly that of a dipole, and more importantly the magnetic field created in a building in activity is not neglectible (especially in a physics lab hosting NMR experiments). When we did the experiment, we measured 25 μT at the bottom of the building and 38 at the top, a 40% difference which an ideal model would translate in a height of several hundreds of kilometres.

The variation with the altitude of the value of $g$ is mentioned in Bohr's legend as a possible way to determine the height of the building (at such level of precision, the variation of $g$ also depends on the density of the building and its foundation [42]). Using a smartphone, one could then build two pendulums, and measure $g$ through the period value (#57) or simply read the value of Earth gravity given by the smartphone's accelerometer (#58). Both solutions are equally non-realistic. Not considering that we are assuming a perfect round Earth and neglecting the effect of neighbouring masses on the value of local $g$, assuming a 1-meter pendulum and a 0.1-second resolution in the period measurement gives an uncertainty of 3.2 km in the height value if using the former method. Using the latter method, our smartphone accelerometer had a slow drift of roughly 0.01 ms$^{-2}$, which corresponds to 3 km of uncertainty. We can safely conclude that for our building $H = 0 \pm 3000$ m.

Still playing with the idea of a slight change in local $g$ with altitude, general relativity tells us that time is not the same at the top and at the bottom of the building: [43] if two smartphone stopwatches are started at the same time at the bottom of the building and one of them is brought up to the top for a given time $t$, say 1 hour, then brought down, a delay $\delta t$ should exist between the two stopwatches:

$$\frac{\delta t}{t} = \frac{\langle g \rangle H}{c^2}$$

with $\langle g \rangle$ the averaged value of $g$, and $c$ the speed of light. Assuming that we are able to detect a 1 ms difference between the two stopwatches on a 1-hour experiment, this would result in an uncertainty of 3000000 km on $H$! This method seems farfetched, but atomic clocks do have the resolution to measure this effect on Earth [43] (with an altitude difference much higher than a building height). Special relativity tells us that when the clock is brought up and down, the speed of displacement $v$ will also change the local time: the correction is given by $v^2/2c$ compared to the effect of general relativity $gH/c^2$. Back of the envelope calculations shows that the effect of velocity is neglectible.

Table 1: List of the 61 methods. The methods are described in the text. A result given as 0 means that the experiment was carried out but the height could not be calculated. When the result is not given and the number is starred (e.g. #6*), we did not test the method. Technical details are provided in appendix A.

| Nb | Method | Results (m) |
|---|---|---|
| 1 | Free fall of the smartphone | 14.0 ± 1.1 |
| 2 | Free fall of an object, using a stopwatch | 16 ± 2 |
| 3 | Free fall of an object, using video analysis | 17.7 ± 0.6 |
| 4 | Free fall of an object, using audio analysis | 18.1 ± 0.3 |
| 5 | End velocity of the free fall of an object, video analysis | 13.9 ± 0.7 |
| 6* | End velocity of the free fall of a speaker, Doppler analysis | * |
| 7* | Distance of landing for an object thrown horizontally | * |
| 8 | Multiple rebounds of a ball, video analysis | 6.5 ± 1.5 |
| 9 | Multiple rebounds of a ball, audio analysis | 6.5 ± 1.5 |
| 10 | Giant pendulum, using a stopwatch | 14.4 ± 0.4 |
| 11 | Giant pendulum, video analysis | 14.7 ± 0.4 |
| 12 | Giant pendulum, using the accelerometer | 0 |
| 13 | Giant pendulum, using the gyroscope | 14.4 ± 0.4 |
| 14 | Giant pendulum, using the magnetometer | 14.1 ± 0.4 |
| 15 | Giant pendulum, using the light sensor | 14.3 ± 0.4 |
| 16 | Giant pendulum, using the proximity sensor | 14.9 ± 0.4 |
| 17* | Giant pendulum, audio analysis | * |
| 18 | Giant torsional pendulum, any sensor | 10.8 ± 0.45 |
| 19* | Relation between centripetal acceleration and angular velocity (on a giant pendulum) | * |
| 20* | Relation between angular velocity and velocity (on a giant pendulum) | * |
| 21 | Thales' method with shadows | 16.8 ± 1.3 |
| 22 | Shadow length and sun elevation from GPS data | 14.8 ± 0.3 |
| 23 | Shadow length and sun elevation at the equinox | 13.8 ± 0.2 |
| 24 | Measuring the angle from eye level to the top with the accelerometer | 15.5 ± 0.4 |
| 25 | Measuring the angle from the bottom to the top with the accelerometer | 40 ± 20 |
| 26 | Measuring the angle of view of an object on the ground from the top with the accelerometer | 15.3 ± 1.0 |
| 27 | Measuring the angle of view of an object on the ground from a picture | 13.6 ± 1.8 |
| 28 | Picture with a scale of the building | 14.8 ± 0.1 |
| 29 | Picture of the building knowing the specifications of the camera | 14.6 ± 0.2 |
| 30 | Picture of an object on the ground from the top knowing the specifications of the camera | 14.9 ± 0.2 |
| 31 | Length of a rope along the facade | 14.5 ± 0.1 |
| 32 | Length of a rope along the façade, using a pulley and the gyroscope | 15.0 ± 0.2 |

| | | | | | |
|---|---|---|---|---|---|
| 33 * | Length of a rope along the façade, using a double integration of the accelerometer data | * | 50 | Decrease of light intensity with distance | 15.8 ± 0.5 |
| 34 | Piling up smartphones along the facade | 14.8 ± 0.2 | 51 | Decrease of Wifi intensity with distance | 23 ± 11 |
| 35 | Number of stairs to the top | 15.0 ± 0.2 | 52 | Decrease of magnetic field intensity with distance | * |
| 36 | Variation of atmospheric pressure | 15.1 ± 0.1 | 53 | Decrease of radioactive intensity with distance | * |
| 37 * | Double integration of the accelerometer during and elevator ride | * | 54 | Decrease of the surface on a picture with distance | 14.7 ± 1 |
| 38 | Altitude difference from the GPS | 8 ± 10 | 55 | Projection of a hair diffraction pattern from the top to the ground | 16.5 ± 1.7 |
| 39 | Sound time of flight, using acoustic stopwatches | 15.0 ± 0.3 | | | |
| 40 | Sound time of flight, using two synchronized audio recordings | 14.5 ± 0.4 | 56 | Projection of a smartphone screen diffraction pattern from the top to the ground | 15.5 ± 0.2 |
| 41 | Sound time of flight, using two audio recordings synchronized by a phone call | 10.9 ± 1.4 | 57 | Variation of gravity between the top and the ground, determined using small pendulums | 0 ± 3200 |
| 42 | Sound time of flight, from the echo | 0 | 58 | Variation of gravity between the top and the ground, determined by the accelerometer | 0 ± 3000 |
| 43 | Sound time of flight, using slow-motion movie | 15.2 ± 1.0 | 59 | Variation of the Earth magnetic field between the top and the ground | 8E5 ± 1E5 |
| 44 | Audio phase shift along the facade of a single frequency | 14.9 ±0.3 | | | |
| 45 | Audio phase difference from the top and the bottom when changing the frequency | 13.5 ± 1.3 | 60 | Variation of gravity between the top and the ground, determined by general relativity time dilatation | 0 ± 3 E9 |
| 46 * | Audio phase shift when moving laterally at the top, a single frequency being emitted at the bottom. | * | 61 * | Phone call to the building's architect | * |
| 47 * | Acoustic interferences at the top created by two speakers at the bottom | * | | | |
| 48 * | Resonance of a tube along the facade | * | | | |
| 49 * | Decrease of sound intensity with distance | * | | | |